\documentclass[12p]{amsart}

\usepackage{graphicx}
\usepackage{amsmath,amsfonts,amssymb}

\usepackage[applemac]{inputenc}
\usepackage{color}

\newcommand{\bx}{\mathbf{x}}
\newcommand{\bu}{\mathbf{u}}
\newcommand{\bd}{\mathbf{d}}
\newcommand{\bq}{\mathbf{q}}
\newcommand{\be}{\mathbf{e}}
\newcommand{\br}{\mathbf{r}}

\newcommand{\bN}{\mathbf{N}}
\newcommand{\bB}{\mathbf{B}}
\newcommand{\bK}{\mathbf{K}}

\newcommand{\bs}[1]{\boldsymbol{#1}} 
\newcommand{\bsigma}{\boldsymbol{\sigma}} 
\newcommand{\bepsilon}{\boldsymbol{\epsilon}}
\newcommand{\bkappa}{\boldsymbol{\kappa}}
\newcommand{\bzeta}{\boldsymbol{\zeta}}

\title[]{On damping created by heterogeneous yielding in the numerical analysis of nonlinear reinforced concrete frame elements}

\date{}

\begin{document}

\maketitle

\renewcommand*{\thefootnote}{\fnsymbol{footnote}}

\begin{center}
Pierre~Jehel\textsuperscript{1,2}\footnote{Corresponding author: pierre.jehel[at]centralesupelec.fr} and R\'egis~Cottereau\textsuperscript{1} \\
\vspace{0.5cm}
$^1$ Laboratoire MSSMat / CNRS-UMR 8579, \'Ecole Centrale Paris, Grande voie des Vignes, 92295 Ch\^atenay-Malabry Cedex, France\\
$^2$ Department of Civil Engineering and Engineering Mechanics, Columbia University, 630 SW Mudd, 500 West 120th Street, New York, NY, 10027, USA \\
\vspace{0.5cm}
\small{March 2$^{\textrm{\tiny{nd}}}$, 2015: Paper accepted for publication in \emph{Computers and Structures}} \\
\small{DOI: 10.1016/j.compstruc.2015.03.001}
\end{center}

\renewcommand*{\thefootnote}{\arabic{footnote}}

\begin{abstract}
In the dynamic analysis of structural engineering systems, it is common practice to introduce damping models to reproduce experimentally observed features. These models, for instance Rayleigh damping, account for the damping sources in the system altogether and often lack physical basis. We report on an alternative path for reproducing damping coming from material nonlinear response through the consideration of the heterogeneous character of material mechanical properties. The parameterization of that heterogeneity is performed through a stochastic model. It is shown that such a variability creates the patterns in the concrete cyclic response that are classically regarded as source of damping.
\end{abstract}


\section{Introduction}
\begin{figure}[htb]
\begin{center}
 \includegraphics[width=7cm]{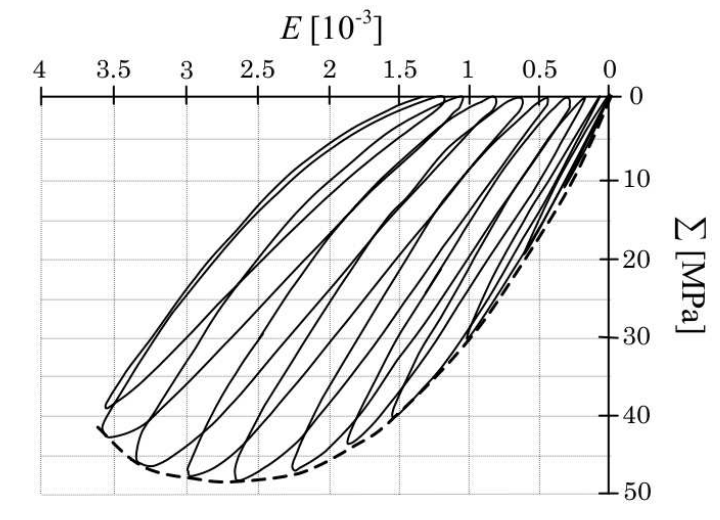}
\caption{Strain-stress concrete experimental response in pseudo-static cyclic uniaxial compressive loading (adapted from~\cite{Ramtani1990}). $\Sigma$ and $E$ are the homogeneous compression stress and strain in the concrete test specimen that are measured in the loading direction. $\Sigma$ and $E$ are spatial mean quantities in the sense that $\Sigma$ is computed as the load in the hydraulic cylinder of the testing machine divided by the area of the specimen cross section, and $E$ is computed as the displacement of the cylinder divided by the length of the concrete specimen.}
\label{fig:EssaiRamtani}
\end{center}
\end{figure}

In the last few decades, a great deal of attention was paid to the comprehension and modeling of damping mechanisms in inelastic time-history analyses (ITHA) of concrete and reinforced concrete (RC) structures~\cite[section 2.4]{PEER/ATC-72-1}. Figure~\ref{fig:EssaiRamtani}, adapted from~\cite{Ramtani1990}, shows the uniaxial cyclic compressive strain-stress ($E$-$\Sigma$) response measured on a concrete test specimen. Throughout this paper, the term ``uniaxial" implies that there is only one loading direction and that the stress, respectively strain, of interest is the normal component of the stress, respectively strain, vector in the loading direction. In other words, when it comes to constitutive relation between stress and strain, the work presented thereafter is developed in a 1D setting. In figure~\ref{fig:EssaiRamtani}, the so-called backbone curve, which is the envelope of the response (dashed line), shows the following phases: (i) an inelastic phase with positive slope ($E \leq 2.7\times\smash{10^{-3}}$ for that particular example, where $E$ is the measured strain), and (ii) an inelastic phase with negative slope before the specimen collapses. For concrete, no elastic phase can really be identified, and hysteresis loops appear in unloading-reloading cycles even for limited strain amplitudes. Other salient features include: (i) a residual deformation after unloading, and (ii) a progressive degradation of the stiffness (slope of the unloading-loading segments). The hysteresis loops are one of the sources of the damping that is observed in free vibration recordings of concrete beams. Other sources include friction at joints~\cite{LesBac2007} or at the concrete-steel interface in reinforced concrete~\cite{DomIbr2012}. These other sources of damping will not be discussed in this paper, where we will concentrate on material damping.

\begin{figure}[htb]
\centering
\includegraphics[width=0.8\textwidth]{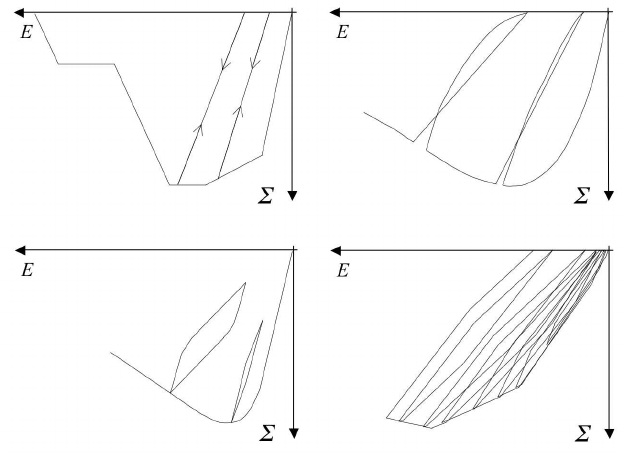}
\caption{Typical strain-stress relations in \ pseudo-static cyclic compressive loading for different models: \cite{Perform3D}~[top left], \cite{Vector2}~[top right], \cite{RagLaBMaz2000}~[bottom left], and~\cite{Jeh-et-al2010}~[bottom right].}
\label{fig:con-mod}
\end{figure}

Most classical uniaxial constitutive models of concrete for numerical simulation do not dissipate any energy in unloading-reloading cycles (see e.g. figure~\ref{fig:con-mod} [top left]). It is then common practice to add a viscous damping model (Rayleigh damping) to the inelastic structural model, to reproduce phenomena that are experimentally observed at the structural level (decreasing amplitude of displacements in free vibration). However, Rayleigh damping is well known to lack physical justification, even when care is taken to avoid generating spurious damping forces~\cite{Charney2008, Hall2006}.

Another class of approach aims at reproducing more precisely the features of Figure~\ref{fig:EssaiRamtani} through elaborate inelastic constitutive relations. Figure~\ref{fig:con-mod} shows typical examples of uniaxial relations found in the literature. The relation described in~\cite{Perform3D} [top left] defines different response phases for different strain intensities, with an additional coefficient to control the loss of stiffness. The constitutive relation described in~\cite{Jeh-et-al2010} [bottom right] comes from a formulation developed in the framework of thermodynamics with internal variables. It reproduces damping features reasonably well, in particular for higher amplitudes, but requires the identification of a rather large number of parameters. These first two types of relations are somehow defined by parts for different loading regimes. They hence require a wider set of parameters and seem to contradict the seemingly smooth transition between regimes observed experimentally. The constitutive relation described in~\cite{Vector2} [top right] is heuristically defined from a database of experiments. It reproduces unloading-reloading hysteresis mechanisms, but lacks a theoretical basis. Finally, the relation described in~\cite{RagLaBMaz2000} [bottom left] is based on a physical model of damage and friction. It manages to dissipate energy in unloading-reloading cycles, but the lack of obvious physical meaning for some parameters can render their identification difficult.

The main purpose of this paper is to present a multi-scale stochastic nonlinear concrete model that can be accommodated in an efficient structural frame element (fiber element), and that participates to the overall structural damping in dynamic loading. In particular, this implies the developed concrete model be capable of representing hysteresis loops in unloading-reloading cycles at macro-scale (the scale where such behavior as in Figure~\ref{fig:EssaiRamtani} can be observed). In this work, this is achieved by the introduction, at an underlying meso-scale, of spatial variability in the parameters. At meso-scale, an elasto-plastic response with linear kinematic hardening and heterogeneous yield stress is considered. This choice is mainly driven by its simplicity and its relevance is illustrated in the numerical applications.

The main issue with modeling the heterogeneity of the yield stress lies in the parameterization. On the one hand, local information on the heterogeneity of concrete is available at a scale that we wish to avoid (because of the associated computational costs). On the other hand, identification becomes extremely difficult when very fine models are considered. We therefore choose to model the heterogeneity of the yield stress through a stochastic model. Hence, only three parameters control that heterogeneity: a mean value, a variance, and a correlation length. The choice of parameterizing the fluctuating field of constitutive parameters by statistical quantities means that there might be fluctuations in the quantities of interest measured for different realizations of the random model. However, as will become apparent in the examples, some sort of homogenization comes in and these fluctuations can rightfully be ignored.

Several authors in the literature have considered random models of fluctuating nonlinear materials~\cite{Frantziskonis1998, Huet1997, Anders2001, RosMat2008}, in particular for concrete~\cite{Huh2001, ReaFilMae2003, LeeMos2004, Wriggers2006, Stroeven2008, Yang2009, Namikawa2013} or in the context of dynamic analysis~\cite{Schueller1999, Li2004, SteFra2009}. We consider here a modeling framework that is a combination of ingredients found in several previous papers~\cite{Liu1986, Brenner1995, Huh2001, Namikawa2013}, with a fluctuating yield stress modeled as a random field with non-zero correlation length. However, the objective in these papers was to assess the influence of parameter uncertainty on some quantity of interest. An objective with the current paper is to observe the effect of randomness at a meso-scale on the nonlinear stress-strain relation at macro-scale. The work herein presented should therefore be seen as an innovative proposal for parameterization of a nonlinear stress-strain relation.

In Section~\ref{sec:beam_model}, we recall the theoretical formulation of the inelastic beam model that will be used throughout this paper. The stochastic multi-scale constitutive relation developed to represent concrete cyclic behavior in reinforced concrete frame elements is introduced in section~\ref{sec:concr-law}. Concrete behavior at macro-scale is retrieved from the description of a meso-scale where elasto-plastic response with linear kinematic hardening and spatially variable yield stress is assumed. In particular, we emphasize in sections~\ref{sec:yield-str-RF} and~\ref{sec:num-RF} the heterogeneity of the yield stress and the parameterization of that heterogeneity through a random model. In section~\ref{sec:VanishingCorrelationLength}, we report on the limiting case of vanishing correlation length and monotonic loading, for which several results can be derived analytically. Section~\ref{sec:NumApplications} presents numerical applications of the model in the context of dynamic structural analysis of reinforced concrete frame elements.

%%%%%%%%%%%%%%%%%%%%%%%%%%%%%%%%%%%%%
\section{2D continuum Euler-Bernoulli inelastic beam}\label{sec:beam_model}

Classical displacement-based formulation has been retained here although other mixed formulations can in certain cases show better performances~\cite{Tay-et-al2003}. For the sake of conciseness, we present the beam element in the 2D case, extension to 3D is straightforward.

%------------------------------------------------------------------------------------------------------------------------
\subsection{Euler-Bernoulli kinematics}

We define the continuum beam $\mathcal{B} = \{ \bx \in \mathbb{R}^3 \vert x_1 \in [0 , L] ; \ x_2 \in [-h\slash2 , h\slash2] ; \ x_3 \in [-w\slash2 , w\slash2] \}$. Such a beam has length $L$ and uniform rectangular cross-section $\mathcal{S}$ of size $w \times h$. We consider an orthonormal basis $(\mathbf{i}_1 , \mathbf{i}_2 , \mathbf{i}_3)$ of $\mathbb{R}^3$, so that any material point in space $\bx = \sum_{i=1}^3 x_i\mathbf{i}_i$. In the 2D setting adopted here, Euler-Bernoulli kinematics can be written at any point $\bx \in \mathcal{B}$ and at any time $t \in [0, T]$ as
\begin{equation} \label{eq:beam-kinematics}
 \mathbf{u}(\bx,t) = \left(
 \begin{array}{l}
  u_1(\bx,t) = u_1^S(x_1,t) - x_2 \theta_3^S(x_1,t) \\
  u_2(\bx,t) = u_2^S(x_1,t)
 \end{array} \right)
\end{equation}
with
\begin{equation*}
 \theta_3^S(x_1,t) = \frac{\partial u_2^S(x_1,t)}{\partial x_1} \ .
\end{equation*}
$u_1$ and $u_2$ are the longitudinal and transversal components of the displacement vector $\bu$ at any point in the beam. $u_1^S$ and $u_2^S$ are rigid body translations and $\theta_3^S$ is rigid body rotation of section $\mathcal{S}$ at position $x_1$ along the beam axis. Thus, $u_1^S$, $u_2^S$ and $\theta_3^S$ only depend on $x_1$.

For small transformations, strain tensor reads $\mathbf{E} = \frac{1}{2} \left(\mathbf{D}(\bu) + \mathbf{D}^T(\bu) \right)$, where $\mathbf{D}(\cdot) = \sum_{i=1}^3 \frac{\partial \cdot}{\partial x_i} \otimes \mathbf{i}_i$, with $\otimes$ the tensor product and $\cdot^T$ the transpose operation. Then, defining the axial strain $\epsilon^S = \partial u_1^S \slash \partial x_1$ and the curvature $\chi^S = \partial^2 u_2^S \slash \partial x_1^2$, it comes:
\begin{equation}
 \mathbf{E}(\bx,t) = E(\bx,t) \ \mathbf{i}_1 \otimes \mathbf{i}_1
\end{equation}
with
\begin{equation}\label{eq:axial-strain}
 E(\bx,t) = \epsilon^S(x_1,t) - x_2 \chi^S(x_1,t) \ .
\end{equation}

%------------------------------------------------------------------------------------------------------------------------
\subsection{Variational formulation}

Suppose $\mathcal{B}$ is loaded with forces per unit length of beam $\mathbf{b}$ and concentrated forces applied at beam ends. A variational form of the problem of finding $\bu$ such that beam equilibrium is satisfied is: find $\bu$ such that
\begin{equation} \label{eq:virt-work}
 0 = \int_L \left\{ \int_{\mathcal{S}} \left( \delta \epsilon^S - x_2 \delta \chi^S \right) \Sigma \ d\mathcal{S} \right\} dx_1 - \int_L \delta\bu \cdot \mathbf{b} \ dx_1 - \delta \Pi_{bc} \ ,
\end{equation}
where $\delta\bu$ is any kinematically admissible displacement field $\Sigma = \bs{\Sigma} \cdot \mathbf{i}_1 \otimes \mathbf{i}_1$ with $\bs{\Sigma}$ the stress tensor, $\cdot$ a matrix product here, and $\delta \Pi_{bc}$ the potential for the forces at beam ends.

Introducing the normal and bending forces in the beam cross-sections as
\begin{equation}\label{eq:int-forces}
 N = \int_S \ \Sigma \ d\mathcal{S} \quad \textrm{and} \quad M = - \int_S x_2 \Sigma \ d\mathcal{S} \ ,
\end{equation}
equation~\eqref{eq:virt-work} can then be rewritten as
\begin{equation} \label{eq:int-ene}
 0 = \int_L \delta \be^{\mathcal{S}} \cdot \bq \ dx_1 - \int_L \delta\bu \cdot \mathbf{b} \ dx_1 - \delta \Pi_{bc} \ ,
\end{equation}
where $\bq = \left(N , M \right)^T$ and $\be^{\mathcal{S}} =\left(\epsilon^{\mathcal{S}} , \chi^{\mathcal{S}}\right)^T$.

%------------------------------------------------------------------------------------------------------------------------
\subsection{Inelastic constitutive behavior}

Cross-section inelastic constitutive response $\bq(\be^{\mathcal{S}}; t)$ is thereafter represented using uniaxial material constitutive response $\Sigma(E;\bx, t)$ integrated over the cross-section, rather than using a direct relation between section displacements and forces. This approach leads to what is often referred to as fiber beam element.

With $\Delta$ denoting an increment of some quantity, we introduce the tangent modulus $D$ as $\Delta\Sigma = D \times \Delta E$, that is, from equation~(\ref{eq:axial-strain}),
\begin{equation} \label{eq:macroD}
 \Delta\Sigma(\bx,t) = D(\bx,t) \left( \Delta\epsilon^S(x_1,t) - x_2 \ \Delta\chi^S(x_1,t) \right) \ .
\end{equation}
Introducing relation~(\ref{eq:macroD}) in~(\ref{eq:int-forces}), beam section inelastic constitutive equation reads $\Delta\bq = \mathbf{K}^{\mathcal{S}} \Delta\be^{\mathcal{S}}$ with tangent stiffness matrix
\begin{equation}
\mathbf{K}^{\mathcal{S}}(x_1,t) =
\left[ \int_{\mathcal{S}} 
 \begin{pmatrix}
  1 \\
  -x_2
 \end{pmatrix} D(\bx,t)
 \begin{pmatrix}
  1 & -x_2
 \end{pmatrix} d\mathcal{S} \right] \ .
\end{equation}

%------------------------------------------------------------------------------------------------------------------------
\subsection{Numerical implementation (structural level)}

The finite element method is used to approximate the displacement fields. Classically, we have for each element $\bu(\bx,t) = \bN(\bx) \bd(t)$ and $\be^{\mathcal{S}}(\bx,t) = \bB(\bx) \bd(t)$, where the vector $\bd$ gathers the displacements $u_1^\mathcal{S}, u_2^\mathcal{S} ,\theta_3^\mathcal{S}$ at the element nodes, and matrices $\bN$ and $\bB$ gather the classical shape functions for Euler-Bernoulli kinematics. Choosing, in equation~\eqref{eq:int-ene}, $\delta\bu = \bN \delta\bd$ and $\delta\be^{\mathcal{S}} = \bB \delta\bd$, and then linearizing the resulting relation, we have
\begin{equation}
 \bK^{(k)}_n \Delta\bd_n^{(k)} = \br^{(k)}_n \ ,
\end{equation}
where $\bK = \int_L \bB^T \bK^{\mathcal{S}} \bB dx_1$ and $\br = \mathbf{f} - \int_L \bB^T \bq dx_1$. $\mathbf{f}$ is the vector of nodal forces calculated from $\mathbf{b}$ and from any concentrated force applied at a beam ends. Subscript $n$ refers to any time step in the loading history; superscript $k$ refers to any Newton-Raphson iteration. 

For any integrable function $g$, line integrals are numerically approximated as $\int_L g(x) dx \approx \sum_{l=1}^{N_l} g(x_l) W_l$ where subscript $l$ refers to a quadrature point and $W_l$ denotes quadrature weight and length. Section integrals are estimated as $\int_{\mathcal{S}_l} g(\bx_l) d\mathcal{S}_l \approx \sum_{F=1}^{N_F} A^F g(\bx^F_l)$, where $A^F$ is the section area of the so-called fiber $F$ and $\bx^F_l$ is the position of the fiber centroid in the control section $\mathcal{S}_l$ at quadrature point $l$.

%%%%%%%%%%%%%%%%%%%%%%%%%%%%%%%%%%%%%
\section{Multi-scale uniaxial cyclic model for concrete}\label{sec:concr-law}

In this section, we present the core objective of the paper, which is a nonlinear uniaxial constitutive model capable of representing salient features of concrete compressive response in cyclic loading. It is based on a simple local (meso-scale) elasto-plastic constitutive relation, for which the yield stress is modeled as a random field. The spatial fluctuations of the yield stress induce at macro-scale constitutive relation $\Sigma(E)$ which resembles that encountered experimentally. Again, as already stated in the introduction, the idea of considering an elasto-plastic constitutive relation with fluctuating yield stress is not novel {\it per se}~\cite{Liu1986, Brenner1995, Huh2001, Namikawa2013}. It has been proposed to assess effects of uncertain parameters on model outputs of interest for engineering practice while our aim here is to stress on the fact that this can be seen as a way of parameterizing material nonlinear constitutive relations. In particular, it will be shown that, in some circumstances, even if randomness is present at meso-scale, our model can predict non-random outputs.

%------------------------------------------------------------------------------------------------------------------------
\subsection{Meso- and macro-scale modeling of concrete}
\begin{figure}[htb]
\begin{center}
 \includegraphics[width=0.4\textwidth]{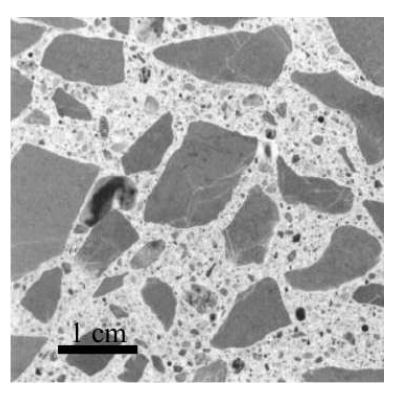}
\caption{Polished concrete section where the two phases human eyes can see are represented: aggregates (crushed gravel and sand) and cement paste in-between.} 
\label{fig:phases}
\end{center}
\end{figure}

Concrete is a heterogeneous material (see figure~\ref{fig:phases}). Two scales are classically considered for its modeling: (i) a micro-scale at which each phase (aggregate, concrete, cement paste) is clearly identified and modeled with its own constitutive relation; and (ii) a macro-scale at which concrete is considered as homogeneous. The macro-scale is the relevant scale for structural engineering applications but the behavior at that scale is strongly influenced by phenomena occurring at the micro-scale. In particular, the geometry of the phases is important as it controls in a large part local concentrations of stresses. As pointed out in the introduction to this paper, formulating and implementing concrete constitutive laws at the macro-scale can then turn out to be challenging, even in the uniaxial case, and especially when it comes to accounting for material energy dissipation sources. We follow here another path, considering a meso-scale at which the parameters of the constitutive relation are assumed to vary continuously. This scale is intermediary between the macro-scale, at which the parameters are homogeneous, and the micro-scale, at which the parameters are discontinuous.

For practical implementation, the heterogeneity will be conveyed in our model by the fluctuations of a random field $\mathfrak{p}(\bx,\theta)$, where $\theta$ represents randomness. Consistently with the fiber beam formulation presented in the previous section, we consider a mesh of fibers $F$ spanning beam cross-sections $\mathcal{S}$. These fibers have a centroid located at position $\bx_l^F$ and a cross-section denoted by $\mathcal{R}$. In the spirit of strain-controlled tests on concrete specimens (see figure~\ref{fig:EssaiRamtani} along with its caption), strain $E$ is assumed homogeneous over $\mathcal{R}$:
\begin{equation} \label{eq:meso-strain}
 \epsilon(\bx,t) = E(\bx_l^F,t) \quad \forall \bx \in \mathcal{R} \ ,
\end{equation}
and stress at macro-scale $\Sigma$ is computed as the spatial mean stress over $\mathcal{R}$:
\begin{equation} \label{eq:mac-str}
 \Sigma(E;\theta) = \frac{1}{|\mathcal{R}|} \int_{\mathcal{R}} \sigma(\epsilon;\theta) d\mathcal{R} \ ,
\end{equation}
where $|\mathcal{R}|$ denotes the area of the fiber section $\mathcal{R}$, $\sigma$ and $\epsilon$ the stress and strain at meso-scale in the uniaxial case. That is, analogously to $E$ and $\Sigma$, $\epsilon = \bepsilon \cdot \mathbf{i}_1\otimes\mathbf{i}_1$ and $\sigma = \bsigma \cdot \mathbf{i}_1\otimes\mathbf{i}_1$, where $\bepsilon$ and $\bsigma$ are the strain and stress tensors at meso-scale such that the constitutive model is considered in a 1D setting. Besides, we enhance the fact that $\Sigma$ is computed as the spatial mean of $\sigma(\bx)$ and not as the sample mean of $\sigma(\theta)$. Local constitutive relation presented in section~\ref{sec:beh-law-con} below governs the relation between heterogeneous stress field $\sigma(\bx,t)$ and strain field $\epsilon(\bx,t)$.

Then, we introduce the tangent modulus at meso-scale $\mathfrak{D}$ as $\Delta \sigma = \mathfrak{D} \times \Delta \epsilon$. According to equations~(\ref{eq:macroD}), (\ref{eq:meso-strain}) and~(\ref{eq:mac-str}), we have the tangent modulus at macro-scale
\begin{equation} \label{eq:mac-mod}
 D(\bx,t;\theta)=\frac{1}{|\mathcal{R}|} \int_{\mathcal{R}} \mathfrak{D}(\bx,t;\theta) d\mathcal{R} \ .
\end{equation}
We will see in the examples below that, for a wide range of relative correlation length and size of the section $\mathcal{R}$, $\Sigma$ and $D$ do not depend on $\theta$. In that case, even though the meso-scale model of concrete is stochastic, the resulting macro-scale model is deterministic, and independent of the actual realization of the local parameters that is being considered.
 
Finally, we recall that concrete specimens exhibit quasi-brittle behavior in tension with tensile strength generally 10 times smaller than compressive strength. Besides, we point out here that it is at macro-scale that these notions of compression ($\Sigma \leq 0$) and tension ($\Sigma > 0$) are relevant for the modeling presented in this work. 

%------------------------------------------------------------------------------------------------------------------------
\subsection{Model of the uniaxial cyclic behavior at meso-scale}\label{sec:beh-law-con}

\begin{figure}[htb]
\begin{center}
\includegraphics[width=0.8\textwidth]{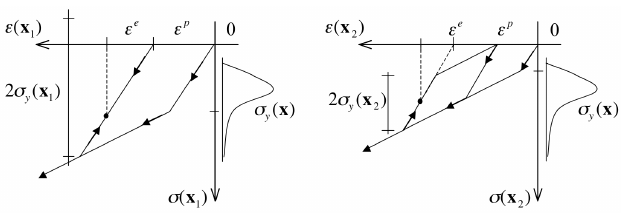}
\caption{Compressive cyclic behavior at meso-scale. The yield stress $\sigma_y(\bx)$ fluctuates over the concrete section $\mathcal{R}$: two local responses at two distinct material points $\bx_1$ [left] and $\bx_2$ [right] are represented in the figure. During elastic loading/unloading, the slope is $C$; in yielding phases, the slope is $\frac{CH}{C+H} < C$. With this type of model first yielding occurs once $\sigma = \sigma_y$ and the elastic domain keeps constant amplitude $2 \sigma_y$. For low values of $\sigma_y$, yielding can be observed in compression both during loading and unloading [right].}
\label{fig:point-law-concrete}
\end{center}
\end{figure}

We now concentrate on the local uniaxial cyclic inelastic constitutive relation that will be considered in this paper at meso-scale. We define it to be a simple elasto-plastic model with linear kinematic hardening, as illustrated in Figure~\ref{fig:point-law-concrete}. We provide here, in the setting of computational inelasticity~\cite{SimHug1998, Ibrahimbegovic2009}, the assumptions and resulting equations corresponding to this relation:
\begin{itemize}
 \item[(i)] The total deformation $\epsilon$ is split into elastic ($\epsilon^e$) and plastic ($\epsilon^p)$ parts:
 \begin{equation}
  \epsilon = \epsilon^e + \epsilon^p \ .
 \end{equation}
 \item[(ii)] The following state equation holds (upper dot denotes derivative with respect to time):
 \begin{equation}
  \dot{\sigma} = C\dot{\epsilon}^e \ ,
 \end{equation}
where $C$ is the elastic modulus.
\item[(iii)] We impose that the stress $\sigma$ corrected by $\alpha$, the so-called back stress due to kinematic hardening, satisfies yielding criterion
\begin{equation} \label{eq:loc-crit}
 \phi^p = |\sigma+\alpha| - \sigma_y \leq 0 \ ,
\end{equation}
where $\sigma_y \geq 0$ is the yield stress. As yielding function $\phi^p(\sigma,\alpha)$ is negative, the material is elastic; otherwise, plasticity is activated and the material state evolves such that the condition $\phi^p(\sigma,\alpha)=0$ is satisfied.
 \item[(iv)] A change in $\epsilon^p$ can only take place if $\phi^p = 0$ and yielding occurs in the direction of $\sigma+\alpha$, with a constant rate $\dot{\gamma}^p \geq 0$:
 \begin{equation}
  \dot{\epsilon}^p = \left\{
  \begin{array}{ll}
   \dot{\gamma}^p \mathrm{sign}(\sigma+\alpha) & \quad \textrm{if} \ \phi^p(\sigma,\alpha) = 0 \\
   0 & \quad \textrm{otherwise}
  \end{array} \right. \ .
 \end{equation}
 $\dot{\gamma}^p$ is the so-called plastic multiplier.
 \item[(v)] With $H$ the kinematic hardening modulus, the evolution of $\alpha$ is defined as:
 \begin{equation}
  \dot{\alpha} = -H \dot{\epsilon}^p = -\dot{\gamma}^p H \mathrm{sign}(\sigma+\alpha) \ .
 \end{equation} 
\end{itemize}

Accordingly, admissible stresses $\sigma$ and $\alpha$ remain in the set $\mathcal{K}^e=\{ (\sigma, \alpha) \ | \ \phi^p \leq 0 \}$ and two kinds of evolutions are possible:
\begin{itemize}
 \item[(i)] If $(\sigma, \alpha) \in \bar{\mathcal{K}}^e = \{ (\sigma, \alpha) \ | \ \phi^p < 0 \}$, the response is elastic:
 \begin{equation}\label{eq:mod-elas}
  \dot{\epsilon}^p = 0 \qquad \Rightarrow \qquad \dot{\sigma} = C \dot{\epsilon} \ .
 \end{equation}
 \item[(ii)] If $(\sigma, \alpha) \in \partial\mathcal{K}^e = \{ (\sigma, \alpha) \ | \ \phi^p = 0 \}$, any evolution is possible only if $\dot{\phi}^p=0$:
 \begin{equation}\label{eq:mod-plas}
  \frac{\partial \phi^p}{\partial \sigma} \dot{\sigma} + \frac{\partial \phi^p}{\partial \alpha} \dot{\alpha} =0 \ \Rightarrow \ \dot{\gamma}^p = \frac{C \mathrm{sign}(\sigma+\alpha)\dot{\epsilon}}{C+H} \ \Rightarrow \ \dot{\sigma} = \frac{CH}{C+H}\dot{\epsilon} \ .
 \end{equation}
\end{itemize}

It is then possible, from equations~(\ref{eq:mod-elas}) and~(\ref{eq:mod-plas}), to give the expression of the tangent modulus $\mathfrak{D}$:
\begin{equation} \label{eq:tang-mod}
 \dot{\sigma} = \mathfrak{D} \dot{\epsilon} \quad \textrm{with} \quad \mathfrak{D} = \left\{
 \begin{array}{ll}
  C \quad & \textrm{if} \ (\sigma, \alpha) \in \bar{\mathcal{K}}^e \\
  \frac{CH}{C+H} \quad & \textrm{if} \ (\sigma, \alpha) \in \partial \mathcal{K}^e
 \end{array} \right. \ .
\end{equation}

It should be reminded at this point that, as mentioned in the introduction and illustrated in Figure~\ref{fig:point-law-concrete}, the yield stress is assumed heterogeneous. The relation presented here is therefore defined between stress and strain in each point in space with a different yield stress.

%------------------------------------------------------------------------------------------------------------------------
\subsection{Description of the yield stress random field} \label{sec:yield-str-RF}

In this section, we describe the choice that is made for the modeling of the heterogeneous yield stress: the yield stress is represented by a 2D log-normal homogeneous random field over the concrete area $\mathcal{R}$. We note here that, to the best of our knowledge, there currently exists no experimental dataset of local stress-strain uniaxial concrete responses recorded at many points over a concrete area. Here, the choice of using random fields to convey heterogeneity of the yield stress is mainly motivated by the effectiveness of the method. We hope that this proposed interpretation of concrete meso-structure will foster interaction between numerical and material scientists and help designing experimental investigations that would eventually support or invalidate the numerical model we propose in this paper.

Let us then consider a probability space $(\Theta,\Omega,\Pr)$, where $\Omega$ is a $\sigma$-algebra of elements of $\Theta$ and $\Pr$ is a probability measure. The 2D random field of yield stress is constructed as a nonlinear point-wise transformation~\cite{Grigoriu1998} $\mathfrak{S}_y(\bx,\theta) = \mathfrak{f}(G(\bx;\theta))$ of a homogeneous unit centered Gaussian random field $G(\bx;\theta)$ with given power spectral density (PSD) $S_{GG}(\bkappa)$. The PSD is chosen here as the product of triangle functions with identical properties in the two orthogonal directions of the 2D plane, denoted by the subscript 1 and 2 throughout sections~\ref{sec:yield-str-RF} and~\ref{sec:num-RF}:
\begin{equation} \label{eq:WK}
 S_{GG}(\bkappa) = \frac{1}{\kappa_u^2} \ \Lambda\left(\frac{\kappa_1}{\kappa_u}\right) \ \Lambda\left(\frac{\kappa_2}{\kappa_u}\right) \ ,
\end{equation}
where $\Lambda(\kappa)=1-| \kappa |$ if $| \kappa |\leq1$ and cut-off wave numbers $\kappa_{u,1} = \kappa_{u,2} = \kappa_u$. For wave numbers above the cut-off $\kappa_u$, the spectral density vanishes. In the spatial domain, this PSD corresponds to the following autocorrelation function (the Fourier transform of $S_{GG}(\bkappa)$):
\begin{equation}
 R_{GG}(\bzeta) = \ \mathrm{sinc}^2 \left( \frac{\kappa_u}{2\pi} \zeta_1 \right) \ \mathrm{sinc}^2 \left( \frac{\kappa_u}{2\pi} \zeta_2 \right) \ ,
\end{equation}
where $\mathrm{sinc}(x) = \mathrm{sin}(\pi x) \slash (\pi x)$. The random field $G(\bx;\theta)$ therefore fluctuates over typical lengths $\ell_{c,1} = \ell_{c,2} = \ell_c = 2\pi/\kappa_u$, the so-called correlation length.

The nonlinear point-wise transformation $\mathfrak{f}$ controls the first-order marginal distribution of $\sigma_y$. In particular, it controls the desired expectation $m$ and variance $s^2$ of the yield stress homogeneous random field. In this paper, we choose to consider a log-normal first-order marginal density, to ensure that the realizations are almost-surely and almost everywhere positive, as expected. The nonlinear transformation is then given by:
\begin{equation} \label{eq:trans-field}
 \mathfrak{S}_y(\bx,\theta) = \mathrm{exp}(m_G + s_G \times G(\bx,\theta)) > 0 \ ,
\end{equation}
where
\begin{equation} \label{eq:mG-sG}
 m_G = - \ln \left( \frac{1}{m} \sqrt{1+\frac{s^2}{m^2}} \right) \qquad \textrm{and} \qquad s_G = \sqrt{ \ln\left(1+\frac{s^2}{m^2}\right) } \ .
\end{equation}

Other first-order marginal densities could be considered, for example using the maximum entropy principle~\cite{Shannon1948, Udwadia1989, Soize2000, Cottereau2007, Cottereau2008} or Bayesian identification~\cite{Beck1998, Howson2005}. Also, the PSD function is translated by the nonlinear transformation $\mathfrak{f}$ so that the PSD of the yield stress and of the underlying Gaussian field are different, with possible incompatibilities with the chosen first-order marginal density~\cite{Grigoriu1998, Puig2004, ShiDeo2013}. These important but technical issues go beyond the scope of this paper and will not be further discussed here.

%------------------------------------------------------------------------------------------------------------------------
\begin{figure*}[ht]
  \includegraphics[width=1.0\textwidth]{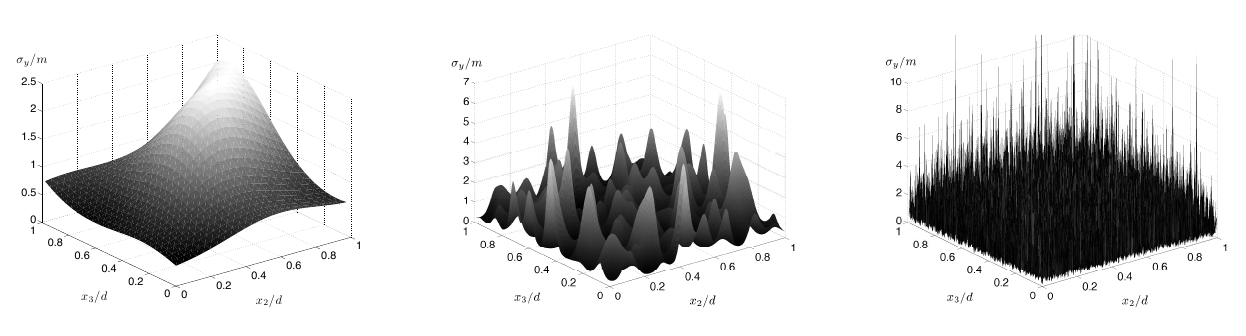}
\caption{Realizations of log-normal random fields over a square of size $d$ for different correlation lengths: [left] $\ell_c \slash d = 1$, [center] $\ell_c \slash d = 0.1$, [right] $\ell_c \slash d = 0$ (white noise). Coordinates $x_2$ and $x_3$, previously introduced in the description of the beam element in section~\ref{sec:beam_model}, are reused here to recall that the random fields are generated to parameter heterogeneous yield stress over beam cross-section areas.}
\label{fig:3RFs}
\end{figure*}

\subsection{Numerical implementation (at each quadrature point in each beam fiber)} \label{sec:num-RF}
At each material point $\bx_l^F$ (quadrature point $l$, beam fiber $F$), relations~(\ref{eq:meso-strain}), (\ref{eq:mac-str}) and~(\ref{eq:mac-mod}) have to be calculated.

On the one hand, Gaussian random field $G$ is digitized using the spectral representation method, in its FFT implementation~\cite{ShiDeo1991, Deodatis1996, Shinozuka1996}. As an illustration, considering a 2D random field with identical properties in orthogonal directions 1 and 2 (see~\cite{Shinozuka1996} for more details):
\begin{multline}
 G(p_1\Delta x,p_2\Delta x;\theta) = \mathrm{Re} \sum_{n_1=0}^{M-1} \sum_{n_2=0}^{M-1} \left( B_{n_1n_2}(\theta) \exp \left(2i\pi \left(\frac{n_1p_1}{M} + \frac{n_2p_2}{M} \right) \right) \right. \\
  \left. + \tilde{B}_{n_1n_2}(\theta) \exp \left(2i\pi \left(\frac{n_1p_1}{M} - \frac{n_2p_2}{M} \right) \right) \right)
\end{multline}
where $i^2=-1$, $(p_1, p_2) \in [0,\ldots,M-1]^2$,
\begin{eqnarray}
&& B_{n_1n_2} = 2\Delta\kappa\sqrt{S_{GG}(n_1\Delta\kappa,n_2\Delta\kappa)} \exp(i \phi_{n_1n_2}(\theta)) \nonumber \\
&& \tilde{B}_{n_1n_2} = 2\Delta\kappa\sqrt{S_{GG}(n_1\Delta\kappa,-n_2\Delta\kappa)} \exp(i \psi_{n_1n_2}(\theta))
\end{eqnarray}
and $\Delta\kappa=\kappa_u \slash N$ ($N \in \mathbb{N}^{\star} \rightarrow \infty$), $M \geq 2N$, $\phi_{n_1n_2}(\theta)$ and $\psi_{n_1n_2}(\theta)$ are independent random phase angles uniformly distributed in $[0, 2\pi]$. The resulting 2D random field is periodic with a two-dimensional period $L_0 \times L_0$ with $L_0 = M \Delta x = 2\pi \slash \Delta\kappa$. Realizations of log-normal random fields with different correlation lengths are shown in figure~\ref{fig:3RFs}.

On the other hand, a generic concrete section $\mathcal{R}$ is built as a square with edge of length $d$ and $\mathcal{R}$ is meshed by a square grid of $N_f^2$ identical squares. Then, the mesh size is $d \slash N_f$ and, for any integrable function $g$, $\int_{\mathcal{R}} g(\bx) d\mathcal{R} \approx \frac{d^2}{N_f^2} \sum_{f=1}^{N_f^2} g(\bx^f)$, where $\bx^f$ is the position of the centroid of the $f$-th mesh over $\mathcal{R}$.

\begin{figure}[htb]
\begin{center}
 \includegraphics[width=0.6\textwidth]{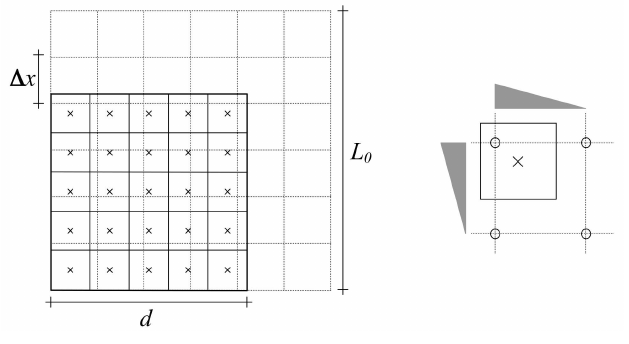}
\caption{On the one hand, the random field is digitized on a square grid (dashed lines) of $(M+1)^2$ points --- here $M=6$ --- spanning an area of size $L_0 \times L_0$, with $L_0 = M \times \Delta x$. On the other hand, generic concrete square section $\mathcal{R}$ has an area of size $d \times d$ ($d \leq L_0$) that is divided into $N_f^2$ identical meshes (plain lines) --- here $N_f=5$. The values of the random field at the positions $\bx^f$ occupied by the centroids of the $N_f^2$ meshes ($\times$) is calculated as the linear interpolation of the four surrounding digitized values of the random field ($\circ$).}
\label{fig:meshes}
\end{center}
\end{figure}

Then, digitized random field $\mathfrak{S}_y$ is mapped onto the $\bx^f$'s over $\mathcal{R}$. To this purpose, we impose $L_0 \geq d$, that is $|\mathcal{R}|$ is smaller or equal to a period of the random field, and mapping is performed according to the following method. First, $N_f$ is calculated as:
\begin{equation} \label{eq:Nf}
 d = \mathrm{Int}\left(\frac{d}{\Delta x}\right) \Delta x + \mathrm{Res} \ \Rightarrow \ N_f = \left\{
 \begin{array}{ll}
  \mathrm{Int}(d \slash \Delta x)       & \mathrm{if} \ \mathrm{Res} = 0 \\
  \mathrm{Int}(d \slash \Delta x) + 1 & \mathrm{otherwise} 
 \end{array}
 \right. \ .
\end{equation}
Then, at the $N_f^2$ points $\bx^f \in \mathcal{R}$, $\mathfrak{S}_y(\bx^f;\theta)$ is calculated as the linear interpolation of the four digitized values of $\mathfrak{S}_y(\bx;\theta)$ in $]\bx^f-\Delta x, \bx^f+\Delta x]^2$, as illustrated in figure~\ref{fig:meshes}.

With the spatially variable yield stress now known at each point $\bx^f$, $f \in [1,..,N_f^2]$ in $\mathcal{R}$, the equations presented in section~\ref{sec:beh-law-con} can be solved to update the variables at meso-scale. This is done numerically at each of the $N_f^2$ positions following classical return-mapping computational procedure~\cite{SimHug1998, Ibrahimbegovic2009}.

Finally, as a transition from compression to tension is detected during global Newton-Raphson iterative process to solve structural equilibrium equations, that is $\Sigma_n^{(k+1)} > 0$ while $\Sigma_n^{(k)} \leq 0$, a local Newton-Raphson precess is implemented to find the strain $E^c$ for which $\Sigma_n^{(k+1)}(E^c)=0$, to update the meso-structure accordingly, and to set $\Sigma_n^{(k+1)}=0$ and $D_n^{(k+1)} = 0$.

Before observing on numerical tests the shape of the stress-strain curves obtained with this model, we turn to the simple case of vanishing correlation length ($\ell_c \rightarrow 0$). The interest of this particular case is that some analytical expressions can be derived, so that discussion is more straightforward. The more general case with finite correlation length will be considered later in Section~\ref{sec:uniaxial_finite_length}.
%------------------------------------------------------------------------------------------------------------------------
\section{A particular case: vanishing correlation length and monotonic loading}\label{sec:VanishingCorrelationLength}

\subsection{Preliminaries}

The case of vanishing correlation length along with uniaxial cyclic loading has been treated in a general setting. Indeed in~\cite{Jeremic2007}, the stress-strain uniaxial response is given as a probability density function ($pdf$) of stress with respect to the time-dependent strain and a second-order exact expression of the $pdf$ evolution is computed solving the Fokker-Planck-Kolmogorov equation that governs the problem. This latter method is valid for monotonic as well as cyclic loading. Hereafter, the validity of the results is limited to monotonic loading, but the problem is cast in a different and simpler mathematical setting that can be solved analytically. These analytical developments shed light on some capabilities of the model introduced in the previous section and that will be retrieved in the more general case of non-zero correlation in Section~\ref{sec:uniaxial_finite_length}.

\subsection{Constitutive response at macro-scale}

Let respectively denote $\mathcal{R}^e$ and $\mathcal{R}^p$ the shares of a fiber cross-section that remain elastic and yield. According to the developments in section~\ref{sec:beh-law-con}: $\mathcal{R}^e(t;\theta) = \{ \bx \in \mathcal{R} \mid \mathfrak{D}(\bx,t;\theta) = C \}$ and $\mathcal{R}^p(t;\theta)=\{ \bx \in \mathcal{R} \mid \mathfrak{D}(\bx,t;\theta) = CH \slash (C+H) \}$. Also, $\mathcal{R}^e \cap \mathcal{R}^p=\emptyset$ and $\mathcal{R} = \mathcal{R}^e \cup \mathcal{R}^p$. Note that $\mathcal{R}^e$ and $\mathcal{R}^p$ are time-dependent because $\mathfrak{D}$ depends on the loading history. We denote by $| \bullet |$ the area of $\bullet$. Then, considering a subset $\mathcal{A}$ of $\mathcal{R}$, we have, $\forall \bx \in \mathcal{R}$, the probability measure $\Pr[\bx \in \mathcal{A}] = |\mathcal{A}| \slash |\mathcal{R}|$.

Using the fact that $|\mathcal{R}|=|\mathcal{R}^e|+|\mathcal{R}^p|$, we first rewrite the tangent modulus at macro-scale in equation~(\ref{eq:mac-mod}) as:
\begin{equation} \label{eq:D-macr-ana}
 D = \frac{1}{|\mathcal{R}|} \left( |\mathcal{R}^e| C + |\mathcal{R}^p| \frac{CH}{C+H} \right) = \frac{C}{C+H} \left( \frac{|\mathcal{R}^e|}{|\mathcal{R}|} C + H \right) \ .
\end{equation}
We now seek an explicit expression for $|\mathcal{R}^e| \slash |\mathcal{R}|$.

First, suppose the state of the material is known at time $t_0$, then we define the trial stresses
\begin{equation} \label{eq:tri-str}
 \sigma^{tr}(\bx,t) = \sigma_0(\bx) + C (\epsilon(t)-\epsilon_0) \quad \textrm{and} \quad \alpha^{tr}(\bx,t) = \alpha_0(\bx) \ ,
\end{equation}
where subscript $0$ refers to time $t_0$. In the particular case of monotonic loading, a necessary and sufficient condition for $\bx$ to be in $\mathcal{R}^p$ at time $t > t_0$ is $\phi^{p,tr}(\bx,t) \geq 0$, that is $\sigma_y(\bx) \leq |\sigma^{tr}(\bx,t) +\alpha_0(\bx)|$ (see equation~\ref{eq:loc-crit}). We then have:
\begin{equation} \label{eq:area-proba}
 |\mathcal{R}^e| \slash |\mathcal{R}| = \Pr[\bx \in \mathcal{R}^e] = 1 - \Pr[\mathfrak{S}_y(\bx) \leq |\sigma^{tr}(\bx,t) + \alpha_0(\bx)|] \ .
\end{equation}

Then, in the particular case of vanishing correlation length, the random variables $\mathfrak{S}_y(\bx)$ are independent and identically distributed over $ \mathcal{R}$. For the log-normal distribution assumption made throughout this work, it means that the cumulative density function of $\mathfrak{S}_y(\bx)$ is, $\forall \bx \in \mathcal{R}$: 
\begin{equation} \label{eq:CDF-sigy}
 \mathcal{F}_{\mathfrak{S}_y(\bx)}(\sigma_y) = \Pr[\mathfrak{S}_y(\bx) \leq \sigma_y] = \frac{1}{2} \left( 1 + \mathrm{erf} \left( \frac{\ln \sigma_y - m_G}{\sqrt{2} \ s_G} \right) \right) \ ,
\end{equation}
where $\mathrm{erf}$ is the so-called error function.

Finally, for the sake of simplicity and without any loss of generality, we assume $\sigma_0 = \alpha_0 = \epsilon_0 = 0$. Accordingly, and using equations~(\ref{eq:tri-str}), along with~\eqref{eq:mG-sG} to replace $m_G$ and $s_G$ by the mean $m$ and standard deviation $s$ of the homogeneous log-normal random field $\mathfrak{S}_y$, it comes:
\begin{equation} \label{eq:prob-phi-neg}
 \frac{|\mathcal{R}^e|}{|\mathcal{R}|} = 1-\mathcal{F}_{\mathfrak{S}_y(\bx)}(|C\epsilon(t)|) = \frac{1}{2} \left( 1 - \mathrm{erf} \left( \frac{\ln \left( \frac{|C\epsilon(t)|}{m} \sqrt{ 1+\frac{s^2}{m^2}} \right)}{\sqrt{ 2\ln \left ( 1+\frac{s^2}{m^2}\right )}} \right) \right) \ .
\end{equation}

\begin{figure}[h]
\begin{center}
 \begin{tabular}{cc}
 \includegraphics[width=0.5\textwidth]{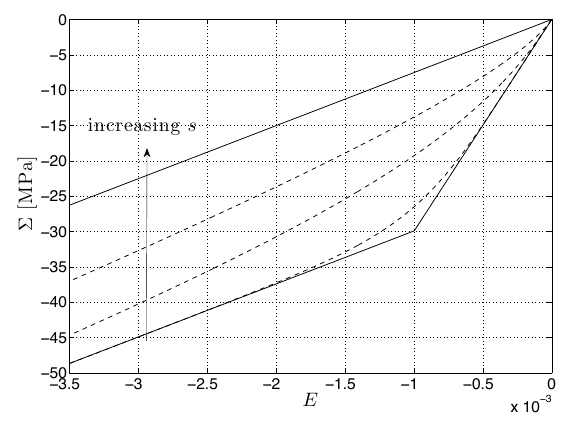} & \includegraphics[width=0.5\textwidth]{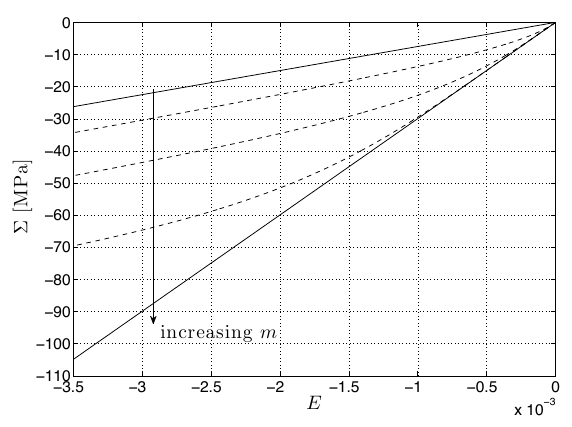}
 \end{tabular}
\caption{Monotonic macro-scale response of the model with vanishing correlation length. Parameters $C=30$~GPa and $H=10$~GPa are used. [top] $m=30$~MPa and $s \slash m = 10^{-4}, 0.4, 2, 10, 10^9$; [bottom] $s=30$~MPa and $s \slash m = 10^{-3}, 0.5, 1, 2.5, 10^3$. Plain curves show asymptotic behaviors as $s \slash m \rightarrow 0$ or $+\infty$.}
\label{fig:axial-loading}
\end{center}
\end{figure}

Equations~(\ref{eq:D-macr-ana}) and~(\ref{eq:prob-phi-neg}) are used to plot figure~\ref{fig:axial-loading} where the response of the model at macro-scale is shown for different sets of mean and variance parameters for the log-normal random yield stress field $\mathfrak{S}_y$.

\subsection{Asymptotic response of the model at macro-scale}

The following asymptotic behaviors can be observed at macro-scale:
\begin{itemize}
 \item[(i)] Suppose $s^2 \slash m^2$ approaches $0$. Then, according to equation~(\ref{eq:prob-phi-neg}), $|\mathcal{R}^e| \slash |\mathcal{R}|$ approaches the Heaviside's function $\mathcal{H}(m-|C E(t)|)$, that is $|\mathcal{R}^e| \slash |\mathcal{R}| = 0$ if $|C E(t)| > m$ and $|\mathcal{R}^e| \slash |\mathcal{R}| =1$ if $|C E(t)| \leq m$. According to equation~(\ref{eq:D-macr-ana}), the model response at macro-scale is then as follows:
\begin{equation}
 \dot{\Sigma}(t) = D(t) \dot{E}(t) \ \textrm{where} \ D = \left\{
 \begin{array}{ll}
  C \quad & \textrm{if} \quad |C E(t)| \leq m \\
  \frac{CH}{C+H} \quad & \textrm{if} \quad |C E(t)| > m
 \end{array} \right. \ .
\end{equation}

 \item[(ii)] If $s^2 \slash m^2 \rightarrow \infty$, then $|\mathcal{R}^e| \slash |\mathcal{R}| \rightarrow 0$ and consequently $\dot{\Sigma}(t) \rightarrow CH \slash (C+H) \dot{E}(t)$.

 \item[(iii)] Now with finite and non-zero $s^2 \slash m^2$:
 \begin{equation} \label{eq:mcr-D-large-eps}
 \dot{\Sigma}(t) = D(t) \dot{E}(t) \ \textrm{where} \ D \rightarrow \left\{
 \begin{array}{ll}
  C \quad & \textrm{if} \quad E(t) \rightarrow 0 \\
  \frac{CH}{C+H} \quad & \textrm{if} \quad E(t) \rightarrow \infty
 \end{array} \right. \ .
\end{equation}
 
\end{itemize}
These asymptotic responses at macro-scale are illustrated in figure~\ref{fig:axial-loading} (plain lines).

%%%%%%%%%%%%%%%%%%%%%%%%%%%%%%%%%%%%%%%%
\section{Numerical applications}\label{sec:NumApplications}
%------------------------------------------------------------------------------------------------------------------------
\subsection{Concrete uniaxial compressive cyclic response at macro-scale}\label{sec:uniaxial_finite_length}

First numerical applications aim at demonstrating the capability of the model introduced above in section~\ref{sec:concr-law} to represent the response of concrete in uniaxial compressive cyclic loading. Five model parameters need to be considered: elastic and hardening moduli $C$ and $H$, along with mean $m$, standard deviation $s$ and correlation length $\ell_c$ used to build realizations of a homogeneous log-normal random field that parameterizes the fluctuations of the yield stress $\sigma_y$ over beam sections.

The effects of $m$, $s$ and $\ell_c$ on the material response at macro-scale will be further investigated below. Right now however, we set:
\begin{itemize}
 \item $C=27.5$~GPa, which corresponds to the elastic modulus measured on specimens made of the concrete actually cast to build the frame element used in the next numerical application (Section~\ref{sec:NA-RC}).
 \item $H=0$ according to both (i) the fact that $H$ controls tangent modulus at macro scale as strain becomes large (see equation~(\ref{eq:mcr-D-large-eps})), and (ii) that we seek a numerical response that ultimately exhibits null tangent modulus in monotonic loading at macro-scale. We anticipate here stressing that the model developed in previous sections is not capable of representing the softening phase as strain increases while stress decreases (non-positive tangent modulus).
\end{itemize}

%%%%%%%%%
\subsubsection{Influence of $\ell_c$ on the macroscopic response.}

\begin{figure*}[htb]
\begin{center}
 \includegraphics[width=1.0\textwidth]{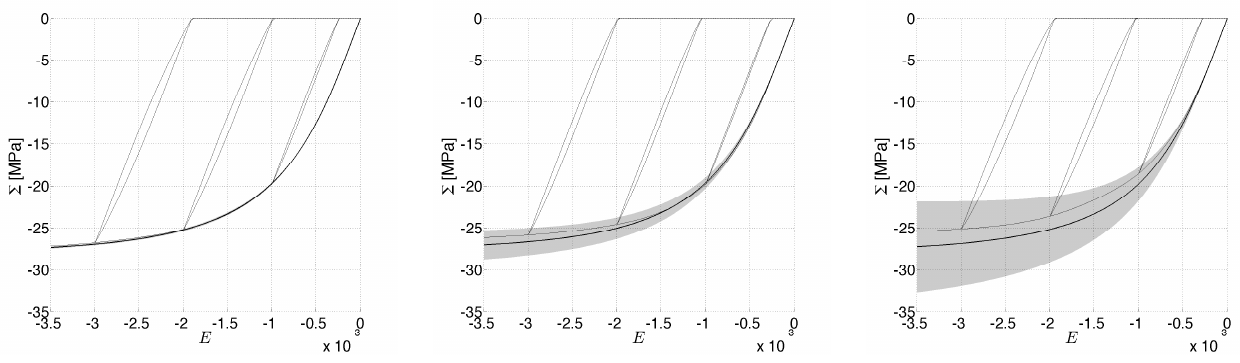}
\caption{Sample mean (thick plain line) plus/minus standard deviation (boundaries of the shaded areas) monotonic response at macro-scale computed from a sample of 100 realizations of the material structure at meso-scale with $m=30$~MPa and $s \slash m=1$ for the log-normal marginal law. Meso-structures are generated with different correlation lengths: [left] $\ell_c \slash d=0.1$, [center] $\ell_c \slash d=0.2$, [right] $\ell_c \slash d=0.4$. Cyclic response for one particular realization of the meso-structure is also shown (thin plain line).}
\label{fig:CompCycLcM}
\end{center}
\end{figure*}

We first illustrate how $\ell_c$ influences the macroscopic response by considering the three following cases: (i) $\ell_c \slash d = 0.1$, (ii) $\ell_c \slash d = 0.2$ and (iii) $\ell_c \slash d = 0.4$. For each of these three cases, we take $N=10$ and $M=64$, that is $\Delta x \slash d = 0.016$ and $N_f = 64$ (see  equation~(\ref{eq:Nf})). We recall that the random field characteristics are taken as identical in both space directions ($\ell_c=\ell_{c,1}=\ell_{c,2}$, $N = N_1 = N_2$, \ldots). Besides, a sample of 100 independent homogeneous log-normal random fields with targeted mean $m = 30$~MPa and coefficient of variation $s \slash m = 1$ for the marginal log-normal law is generated for each case.

Resulting material responses at macro-scale are shown in figure~\ref{fig:CompCycLcM}. A first obvious observation is that model response at macro-scale is much richer than at meso-scale (see figure~\ref{fig:point-law-concrete}). We can then observe that sample mean response (thick line) is not sensitive to the correlation length. However the variability of the macroscopic response from one realization of the meso-structure to another depends on the correlation length: it is almost null for $\ell_c \slash d = 0.1$ while it is enhanced as $\ell_c$ increases. The area $\mathcal{R}$ defined with $\ell_c \slash d = 0.1$, $N=10$ and $M=64$, is statistically representative in the sense that there is almost independence between the random realization of the meso-structure and the response at macro-scale. We remark here that the model is capable of representing variability from one concrete sample to another and that this variability can bring information on the correlations in the meso-structure. Suppose indeed that we had 100 concrete samples and a variability of the responses at macro-scale close to that shown by the grey area in figure~\ref{fig:CompCycLcM} [center] for instance. Then, the meso-structure of the tested concrete would be best represented by the ratio $\ell_c \slash d = 0.2$. Consequently, the sample standard deviation can bring information on the actual correlation length.

Finally, we can notice that the shape of the cyclic response (thin line) represents most of the salient features exhibited experimentally in uniaxial compression test for concrete (remember figure~\ref{fig:EssaiRamtani}). We point out here that strength degradation (softening) along with stiffness degradation (observed experimentally one cycle after another) are not represented by this model. However, a key point for representing material damping is the capability of the model to generate local hysteresis loops in unloading-loading cycles.

\subsubsection{Influence of $m$ and $s$ on the macroscopic response.}

\begin{figure*}[htb]
\begin{center}
 \includegraphics[width=1.0\textwidth]{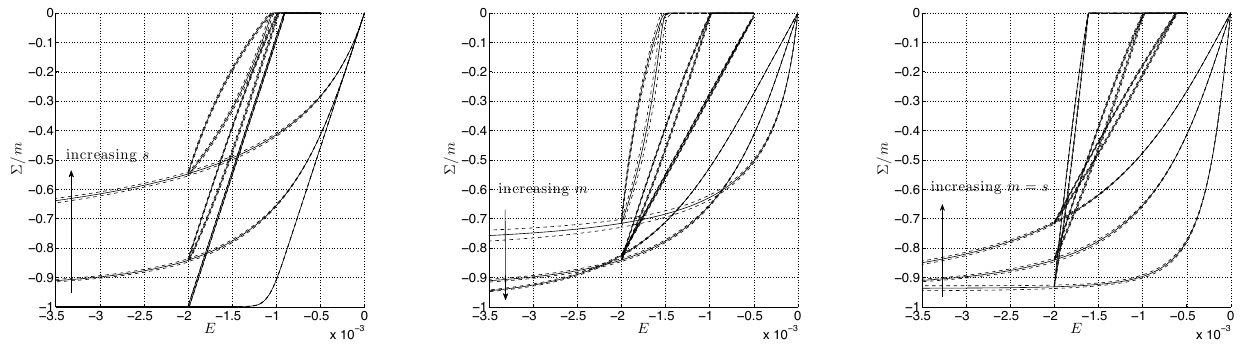}
\caption{Sample mean (plain line) plus/minus standard deviation (dashed lines) response at macro-scale computed from a sample of 100 different realizations of the material structure at meso-scale. Meso-structures are generated with different targeted mean $m$ and coefficients of variation $s \slash m$ for the log-normal marginal law: [left] $m = 30$~MPa and $s \slash m= 0.1, 1, 3$; [center] $s=30$~MPa and $s \slash m=0.6, 1, 3$; [right] $s \slash m = 1$ and $m=10, 30, 50$~MPa.}
\label{fig:CompCycMS}
\end{center}
\end{figure*}

We use here, beside $N=10$ and $M=64$ ($N_f=64$), $\ell_c \slash d = 0.1$ so that little variability is expected to be observed in the model response at macro-scale from one realization of the meso-structure to another (see figure~\ref{fig:CompCycLcM} [left]). Figure~\ref{fig:CompCycMS} shows material response at macro-scale for different sets of mean $m$ and standard deviation $s$ of the log-normal random field that conveys spatial variability in the material structure at meso-scale.

It can be observed that for a small value of $s$, response approaches bi-linear elasto-plastic behavior (actually perfectly plastic because $H$ is set to zero here) where there is almost no hysteresis observed during unloading-loading cycle (figure~\ref{fig:CompCycMS} [left] and [right]). This comes from the fact that, if $s$ approaches $0$, there is almost no spatial variability of the yield stress because it is almost homogeneous over $\mathcal{R}$ and takes values close to its mean $m$; then the response at macro-scale coincides with that at meso-scale (elasto-plasticity with $H=0$ here). This is also in accordance with what was shown already in figure~\ref{fig:axial-loading}.

Responses shown in figure~\ref{fig:CompCycMS} [left] lie in-between this latter extreme case and the other extreme case of $s$ approaching infinity. In this situation, log-normal distribution approaches $0$ all over the positive real semi-line and, consequently, plasticity is activated almost everywhere over $\mathcal{R}$ resulting in a macro-scale response that is perfectly plastic without elastic phase (that is here $\Sigma=0$ for any $E$ because $H=0$).

Also, it is shown in figure~\ref{fig:CompCycMS} that the value of the strain $E$ at which stress $\Sigma$ reaches zero when unloading (residual plastic deformation) is much more sensitive to parameter $m$ than $s$. Furthermore, the thickness of the hysteresis loops obviously depends on the so-called coefficient of variation $s \slash m$ but it is not clear whether it is more sensitive to either of the two parameters. Finally, the variability in the sample of responses at macro-scale increases with $s \slash m$ and is more sensitive to $m$ than $s$, at least as far as the range of values chosen here for both parameters is concerned.

%------------------------------------------------------------------------------------------------------------------------
\subsection{Damping in a reinforced concrete column in free vibration} \label{sec:NA-RC}

We now show how the material model developed in the previous sections can be used to represent the experimental backbone curve of a concrete test specimen in uniaxial loading. Then, we implement this material law in the fiber beam element presented in section~\ref{sec:beam_model} and show how damping is generated in a reinforced concrete (RC) column in free vibration. The observed damping does not result from the addition of damping forces in the balance equation of the RC column but from the hysteresis loops in the concrete material law at macro-scale.

\subsubsection{Geometry of the column and loading.}\label{sec:GeoLoa}

\begin{figure}[htb]
\begin{center}
 \includegraphics[width=0.65\textwidth]{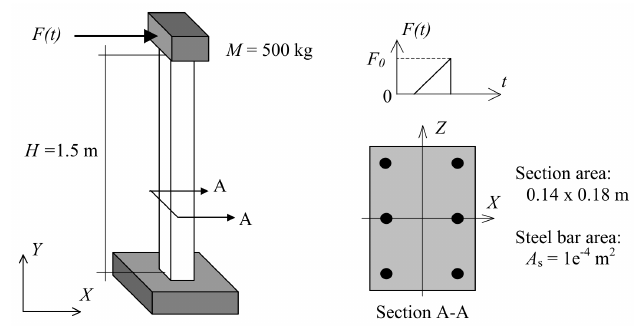}
\caption{Geometry and loading of the column.}
\label{fig:ColCha}
\end{center}
\end{figure}

The column considered here corresponds to the $1$st-floor external column of the ductile ($R=4$) RC frame tested in~\cite{FilLacLam1998a,Lamontagne1997}. The loading is however here different: it consists of a mass $M = 500$~kg imposed step by step and kept constant while the column oscillates in free vibration consequently to a horizontal force $F(t)$. The geometrical and loading characteristics of the column are depicted in figure~\ref{fig:ColCha}.

\subsubsection{Concrete constitutive model.}

\begin{figure}[htb]
\begin{center}
 \includegraphics[width=0.5\textwidth]{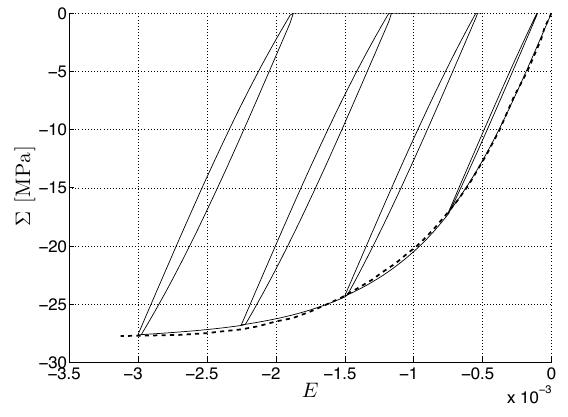}
\caption{Sample mean (---) response at macro-scale obtained numerically from a sample of 2,000 meso-structures, along with backbone curve (- -) recorded during uniaxial test on a specimen of the concrete used to build the RC column of interest here. Targeted mean and standard deviation of the marginal log-normal law are $m = 30.5$~MPa and $s \slash m = 0.943$. No experimental data is available for the cyclic behavior.}
\label{fig:2000}
\end{center}
\end{figure}

In~\cite{Lamontagne1997}, the monotonic uniaxial response (backbone curve) of the concrete cast to build the RC column is detailed. It is used here as the baseline to identify the parameters of the concrete model. According to this report, we set $C = 27.5$~GPa, $H=0$; then we use $N=10$, $M=64$ ($N_f=64$) and $\ell_c \slash d = 0.1$ to ensure a response at macro-scale that is almost independent of the realization of the meso-structure; finally, $m$ and $s$ are identified. Figure~\ref{fig:2000} shows model response macro-scale (plain line) with $m=30.5$~MPa and $s \slash m = 0.943$. Comparing the numerical backbone curve (solid line) and the experimental response (dashed line), this figure illustrates the capability of the developed numerical model to represent actual experimental concrete monotonic response, at least as far as the monotonic behavior is concerned. Note that no experimental data was available for the cyclic behavior.

\subsubsection{Steel cyclic model.}

\begin{figure}[htb]
\begin{center}
\includegraphics[width=0.5\textwidth]{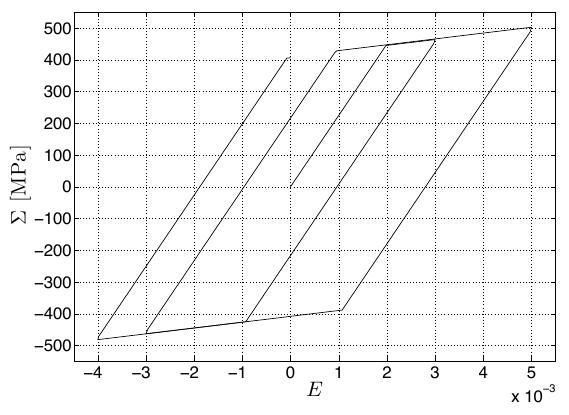}
\caption{Numerical cyclic response of a steel longitudinal rebar used to build the frame. Cyclic behavior has not been observed experimentally.}
\label{fig:LamontagneSteel}
\end{center}
\end{figure}

Young modulus $C_s = 224.6$ GPa, yield stress $\Sigma_y = 438$ MPa and ultimate stress $\Sigma_u = 601$~MPa have been experimentally measured during uniaxial tests on longitudinal steel rebars~\cite{Lamontagne1997}. An elasto-plastic model with kinematic hardening is used to represent steel response in cyclic loading. The model implemented with these latter measured parameters is shown in figure~\ref{fig:LamontagneSteel}.

\subsubsection{Free vibration -- Structural damping.}

\begin{figure}[htb]
\begin{center}
\includegraphics[width=0.7\textwidth]{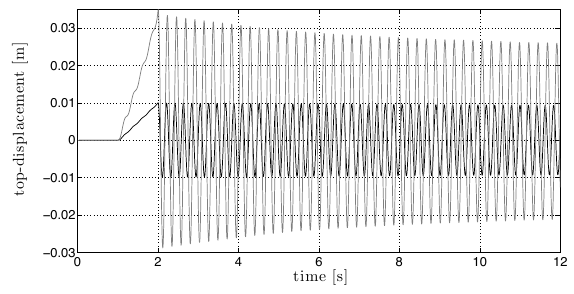}
\caption{Top displacement time history in free vibration for $F_0 = 5$~kN (black) and $F_0 = 15$~kN (grey). Mass $M$ and horizontal forces $F_0$ are applied step-by-step during the first and second seconds, then horizontal force $F$ abruptly drops to zero and the column oscillates in free vibration.}
\label{fig:FV-dis}
\end{center}
\end{figure}

Those concrete and steel uniaxial constitutive models are implemented in the fiber frame element presented in section~\ref{sec:beam_model}. The column is modeled with one frame element with $N_l = 2$ control sections and $N_F = 6$ fibers (actually layers here in the case of a 2D problem). As already mentioned in section~\ref{sec:GeoLoa}, the mass $M = 500$~kg is imposed step by step and kept constant while the column oscillates in free vibration consequently to the horizontal force $F(t)$. The column possibly exhibits nonlinear response while the mass $M$ is applied and while $F(t)$ increases from $0$ to $F_0$. Figure~\ref{fig:FV-dis} shows typical column top-displacement time histories for two different values of $F_0$. It can be observed that damping depends on the amplitude of the oscillations: the column is clearly damped for $F_0=15$~kN (grey curve) while damping is much lower for $F_0=5$~kN (black curve). One can also notice the different vibration periods for both horizontal forces; this is due to the fact that the larger force activates some nonlinear mechanisms in the structure, which leads to an elongation of the structural vibration period. We finally stress again here that there is no damping force added in the dynamic balance equations, such as for instance Rayleigh damping forces: the damping effect shown in figure~\ref{fig:FV-dis} only comes from the hysteresis loops in the concrete response  during unloading-reloading cycles.

We now define what we will refer to as ``viscous-like damping ratio'' and hereafter denote by $\xi^v$. Considering the column top-displacement time history $X_{top}(t)$ in free vibration, we appeal to the so-called log-decrement method to evaluate the modal damping ratio $\xi^v$ (see e.g.~\cite[\S 4.6]{CloPen1975}):
\begin{equation} \label{eq:VLRD}
 \xi^v = \frac{1}{2 \pi N_c}\ln\frac{X_{top}^{peak}(t_{N_1})}{X_{top}^{peak}(t_{N_2})}
\end{equation}
$X_{top}^{peak}(t_{N_1})$ and $X_{top}^{peak}(t_{N_2})$ are the amplitudes of any two peaks separated by $N_c = N_2-N_1$ cycles. It is worth recalling here that this is only valid in case damping is linear viscous, which in our case is not necessarily the case. Indeed, equations~(\ref{eq:VLRD}) comes from the assumption that the envelope of the decaying top-displacement is described as $X_{top}(t)=X_0 e^{-2\xi\pi f t}$ with $f$ the modal frequency. Hence the terms ``viscous-like" to characterize the calculated damping ratios.

\begin{figure}[htb]
\begin{center}
\includegraphics[width=0.5\textwidth]{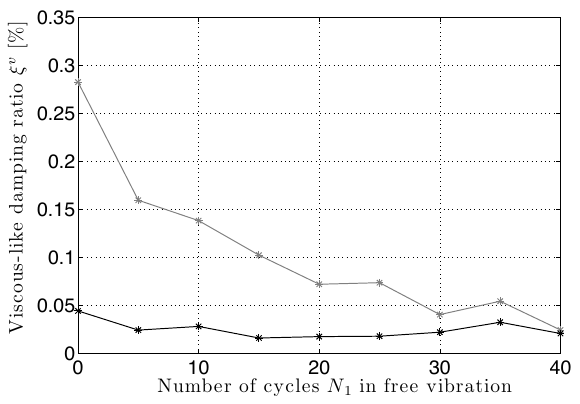}
\caption{Viscous-like damping ratio time history $\xi^v$ for $F_0 = 5$~kN (black) and $F_0 = 15$~kN (grey).}
\label{fig:xi-t}
\end{center}
\end{figure}

\begin{figure}[htb]
\begin{center}
 \begin{tabular}{cc}
  \includegraphics[width=0.5\textwidth]{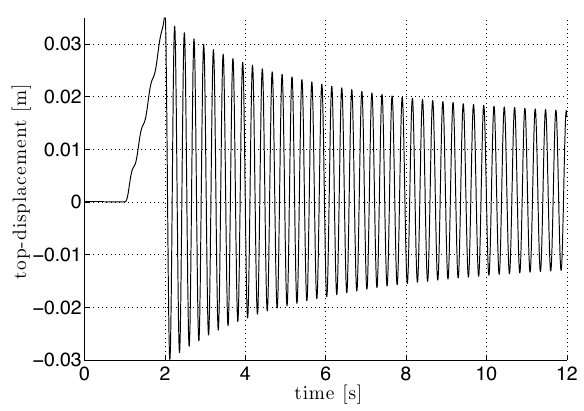} & \includegraphics[width=0.5\textwidth]{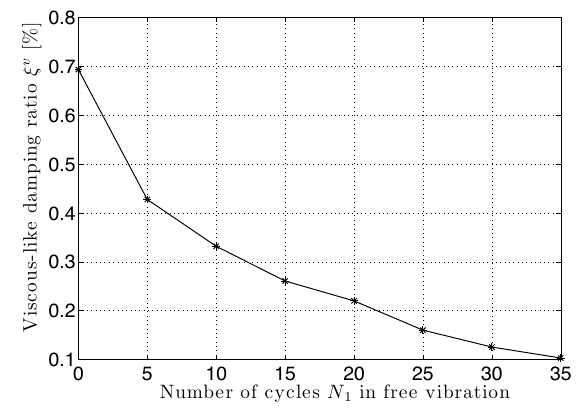}
 \end{tabular}
\caption{[top] Top-displacement time history in free vibration and [bottom] viscous-like damping ratio time history $\xi^v$ for a set of material parameters that is not optimal for the column considered above: $m=20$~MPa, $s \slash m = 10$, $C = 27.5$~GPa and $H=10$~GPa.}
\label{fig:FV-xi-3}
\end{center}
\end{figure}

Based on the top-displacement time histories in figure~\ref{fig:FV-dis}, figure~\ref{fig:xi-t} shows how $\xi^v$ decreases throughout free vibration time history for both values of $F_0$. Viscous-like damping ratios $\xi^v(t_{N_1})$, are computed according to equations~(\ref{eq:VLRD}), with $N_c=5$. Note that such damping ratios depend on the parameters $m$ and $s \slash m$ of the random field along with the hardening parameter $H$ at meso-scale. For the sake of illustration, figure~\ref{fig:FV-xi-3} shows other results for another set of material parameters that is not optimal for representing the monotonic response in compression of the concrete used to build the tested column. The capability of the proposed material model for generating structural damping has been demonstrated and the development of an automatic procedure for identifying the full set of parameters targeting accurate representation of both cyclic concrete response and damping is left for future work. 

%%%%%%%%%%%%%%%%%%%%%%%%%%%%%%%%%%%%%%%%%%
\section{Conclusions}

In this paper, a multi-scale stochastic uniaxial cyclic model suitable for representing most of the salient features of concrete nonlinear response observed in compressive experimental tests has been developed. It is based on the construction of a meso-scale where the response at each material point is elasto-plastic with kinematic hardening and heterogeneous yield stress. This implies that the transition from elastic to plastic regime occurs at a loading level that is different in each material point. Heterogeneity is parameterized by a 2D homogeneous log-normal random field. As a first illustration of the capabilities of the model, some analytical results are derived in the particular case of monotonic loading and vanishing correlation length for the random field. Then, numerical simulations are performed and the effects of the parameters of the random field -- that is the mean $m$, coefficient of variation $s \slash m$ and correlation length $\ell_c$ -- are investigated. It is shown that for small values of the correlation length, material response at macro-scale does not depend on the realization of the random field, showing that the developed model is suitable for an objective representation of the material behavior. Besides, it is shown that the mean $m$ and standard deviation $s$ can be identified so that the monotonic compressive response of an actual concrete test specimen can be accurately represented by the developed model. The developed model however lacks the ingredients for representing both strength and stiffness degradation mechanisms. Finally, the developed material model is implemented in a frame element in the purpose of representing the dynamic response of an actual reinforced concrete column. The numerical analysis of the column in free vibration shows the capability of the developed material model to create patterns classically associated to damping effects. In this simulation, damping does no come from some damping forces added in the dynamic balance equation (e.g. Rayleigh damping) but from the multi-scale stochastic nonlinear model. Although the underlying model is stochastic, the simulations and results shown are the same for any realization of the stochastic model.

The main research prospects lie (i) in the enhancement of the model at meso-scale so that it can represent stiffness and strength degradation mechanisms at macro-scale; (ii) in the precise characterization of the stochastic model based on information from lower scales. This will consist in choosing, based on rational arguments, the type of first-order marginal law and correlation model, as well as the value of the corresponding parameters (mean, variance and correlation length). Although in another context, such an interaction between structural and material scientists has already been appealed for in~\cite{ChaSchPel2007}. Also, these issues could be considered in the context of stochastic micro-meso scale transition~\cite{Soize2008,ArnGha2008,Cot-et-al2011}.

%%%%%%%%%%%%%%%%%%%%%%%%%%
\section*{Acknowledgement}

The first author is supported by a Marie Curie International Outgoing Fellowship within the 7th European Community Framework Programme (proposal No. 275928). The second author, working within the SINAPS@ project, benefited from French state funding managed by the National Research Agency under program RNSR Future Investments bearing reference No. ANR-11-RSNR-0022-04. 



%% References


\bibliographystyle{plain}
\bibliography{2015_CompStruc_PJ-RC}

%% not want to use model3-num-names.bst.





% \end{thebibliography}

%%%%%%%%%%%%%%%%%%%%%%%%%%%%%%%%%%%%%
\end{document}